\documentclass[twocolumn,showpacs,amsmath,amssymb]{revtex4}
\usepackage{txfonts}
\usepackage{bbm}
\usepackage{graphicx}
\usepackage{epsf}

\begin{document}

\title{Pairing Symmetry and Spin Excitations of Iron Selenide Superconductors}

\author
{Tao Zhou$^{1,2}$ and Z. D. Wang$^{1}$}

\affiliation{
$^{1}$Department of Physics and Center of Theoretical and Computational Physics, The University of Hong Kong,
Pokfulam Road, Hong Kong, China\\
$^{2}$College of Science, Nanjing University of Aeronautics and Astronautics, Nanjing 210016, China
}

\date{\today}
\begin{abstract}
We propose an effective single-band model for the newly discovered iron selenide superconductors A$_x$Fe$_{2-y}$Se$_2$ (A=Tl,K,Rb,Cs). Based on this minimum model and the random phase approximation, the $d_{x^2-y^2}$ pairing symmetry is revealed theoretically, which may be understood in the framework of Fermi surface topology. A common origin of superconductivity is elucidated for this compound and other high-T$_c$ materials. The spin excitations at $(\pi,\pi/2)$ in superconducting states are observed, in good agreement with the neutron scattering experiments. The spin resonance is indicated to show up only for the $d$-wave pairing, which provides an additional indication for the $d$-wave pairing symmetry in this family of superconductors.

\end{abstract}
 \maketitle

 \section{Introduction}

Recently, the discovery of a new type of iron-based superconducting (SC) compounds A$_x$Fe$_{2-y}$Se$_2$ (A=Tl,K,Rb,Cs) has attracted much attention ~\cite{guo,mazio,fang}. Both the angle-resolved photoemission spectroscopy (ARPES) experiments~\cite{zhang,mou,zhao,xpw,qian} and the local density approximation (LDA) calculations~\cite{she,yan,cao} have reported that the hole Fermi surface (FS) pockets disappear and only electron ones exist.
Thus
this compound may be heavily electron-doped, which challenges the current physical picture for iron-based SC materials. Due to the absence of the hole-like FS, obviously, the inter-band FS nesting picture~\cite{mazin,yao}, which seems to work well for iron pnictides, is unable to straightforwardly account for the superconductivity of this compound. The pairing symmetry is also an important and critical issue to be addressed, which may be different from that of iron pnictides due to the absence of the hole FS pockets. Actually, the ARPES experiments~\cite{zhang,mou,zhao,xpw} indicated that the FS pockets around the $M=(\pi,\pi)$-point are fully gapped while near the $\Gamma=(0,0)$ point the gap is negligibly small~\cite{zhang,mou}, which is contrast to the $s_{x^2y^2}$-wave symmetry. Up to now, the pairing symmetry is still unclear: both $s$-wave and $d$-wave symmetries have been proposed~\cite{wang,mai,tdas,kot,hu,mazi,sai,yzhou,yu}.

Due to the presence of magnetism in the parent compound, usually the spin excitation is proposed to account for superconductivity in high-Tc materials~\cite{mazin,yao}.
It is natural and important to ask whether this picture still works well in iron selenides. This issue was first studied theoretically based on the five-orbital model. The $d$-wave pairing symmetry, which meditated by the spin dynamics, was proposed~\cite{wang,mai}; while the results of spin susceptibility are different, with the $(\pi,\pi)$-spin excitation proposed in Ref.~\cite{wang} and $(\pi,0.625\pi)$-one proposed in Ref.~\cite{mai}.

Experimentally the inelastic neutron scattering (INS) experiment is one of the most powerful ways to study the spin excitations and can provide us directly the imaginary part of the spin susceptibility in the whole momentum and frequency space. The INS experiments on iron selenides have been reported by several groups. For the insulating parent compound, a $(1/5\pi,3/5\pi)$ spin excitation was reported, indicating the presence of vacancy order and block antiferromagnetic (AF) ground state~\cite{bao,mwang}. For the SC compound, the INS experiments have revealed several channels of spin excitations~\cite{par,frie,miaoyin} around the $(1/5\pi,3/5\pi)$, $(\pi,\pi/2)$, and $(\pi,0)$, respectively. Among them the $(1/5\pi,3/5\pi)$ excitation accounts for the vacancy ordered phase.
The presence of $(\pi,0)$ excitation seems controversial: It was indicated in Refs. ~\cite{par,frie} that no spin excitation at this wave vector was observed, but, on the contrary, the $(\pi,0)$ spin excitation was very recently reported in Ref.~\cite{miaoyin}. It may come from the localized moments in the SC state~\cite{miaoyin}. Interestingly, the $(\pi,\pi/2)$ spin excitation was found to be rather robust in the SC state possibly due to the FS nesting~\cite{par,frie,miaoyin}. Notably,
this excitation is enhanced significantly at the frequency $0.014$ eV, corresponding to the spin resonance at this frequency.
 The resonant spin excitation is a common feature for high-Tc superconductors and thus it
 may be  related to superconductivity intimately, providing an insight
for exploring the mechanism and pairing symmetry of superconductivity.

To study spin excitations and other physical properties theoretically,
it is highly-demanded and desirable to establish a minimum model that is able to capture the essential features of this compound.
Firstly, the vacancy order can be neglected when studying the physics of superconductivity. It has been reported by several
groups that the phase separation occurs in the SC
state, with a vacancy ordered state only existing
in the insulating region~\cite{fchen,ric,zwang,bshen,wli}. Based on the ARPES experiments~\cite{zhang,mou,zhao,xpw,qian}
there is no band folding effect corresponding to the vacancy order wave vector, also seems to propose that the vacancy order does not exist in the SC state. As a result, the vacancy order should not be relevant to superconductivity and need not to be concerned for a minimum model.

From the band calculations for this compound~\cite{she,yan,cao},
it seems that all of the five orbitals hybridize strongly, and thus it was proposed that all
five $d$-orbitals should be considered to construct a model
~\cite{wang,mai,yu}. Note that the five-orbital model is one of the most frequently-used model to describe the iron-pnictide compound~\cite{kur}.
However, this model includes some non-essential parts that may be redundant for superconductivity (especially those bands far away from the FS), and even makes it quite difficult to accurately analyze and understand certain essential physics behind because too many unknown parameters are involved. For example,
if the electron correlation needs to be included, both inter-orbital and intra-orbital interactions should be taken into account;
while the corresponding interaction strengths are difficult to be determined (or adjusted) in order to figure out the relevant physics.
Motivated by this, a minimum two-band model was proposed for iron pnictides~\cite{qhan,rag}, and many physical properties have been understood based on this type of two-band model, especially for those related closely to the low-energy excitations.
When studying the A$_x$Fe$_{2-y}$Se$_2$ materials, a similar two-orbital model was put forward, while the obtained FS size is much larger than that obtained from the band calculation~\cite{ryu}. Intriguingly, as discussed below, we can refine the minimum model of this compound to have one band to capture the essential physics with better results. As seen from the band calculations~\cite{she,yan,cao}, there exist two electron-like FS sheets around the $M$ point. Around the $\Gamma$ point, a small FS pocket may exist, but it is k$_z$ dependent and electron-like. The FSs from ARPES experiments are consistent with the LDA calculations while the electron pockets around the $\Gamma$ point have a very low spectral weight. So it is reasonable to believe that only the FS pockets around M point are essential for constructing a minimum model, while all other bands that do not cross the Fermi energy may be neglected.  Taking into account that one unit cell consists of two iron ions plus the band folding effect, only one kind of FS pocket around the $X=(\pi,0)$ or its symmetric points is actually relevant in the unfolded BZ.

In this paper, motivated by the above considerations, we propose a single-band tight-banding model as a minimum one for the SC A$_x$Fe$_{2-y}$Se$_2$. Our model could have a band dispersion crossing the FS, being qualitatively consistent with that of the five-orbital model~\cite{yu,jxzhu} fitted from the LDA band calculations and the ARPES experiments for the A$_x$Fe$_{2-y}$Se$_2$ materials~\cite{zhang,mou,zhao,xpw,qian}. Thus the present model may serve as an effective one for describing the low energy physics of the A$_x$Fe$_{2-y}$Se$_2$. The normal state spin susceptibility is calculated and then analyzed based on the FS topology. The SC gap is also calculated self-consistently, with the robust $d$-wave SC pairing being revealed. The $d$-wave symmetry can well be understood by the spin fluctuation picture and the fermiology theory. The SC state spin excitations are investigated further with the $d$-wave symmetry. The resonant spin excitation at $(\pi,\pi/2)$ is observed, which is consistent with recent INS experiments and thus provides a certain kind of  support for the $d$-wave symmetry in A$_x$Fe$_{2-y}$Se$_2$ SC materials.

The paper is organized as follows. In Sec. II, we introduce
the model and work out the formalism. In Sec. III, we
perform numerical calculations and interpret the obtained results.
The band structure, pairing symmetry and spin excitations are elaborated, respectively.
Finally, we give a brief summary in Sec. IV.

\section{model and formalism}
We start from a $t-J$ type model including the tight-banding term and the local spin interaction, which reads
\begin{equation}
H=\sum_{{\bf k}\sigma}\varepsilon_{\bf k}n_{{\bf k},\sigma}+H_J,
\end{equation}
where $\varepsilon_{\bf k}$ is phenomenologically taken as the single band tight-banding approximation: $\varepsilon_{\bf k}=-2t(\cos k_x+\cos k_y)-4t^{\prime}\cos k_x\cos k_y-\mu$. $H_J$ is the local spin interaction.
From the first principle calculation in Ref.~\cite{cao}, we assume the spin interaction
to include the nearest-neighbor and the next-nearest-neighbor interactions~\cite{cao}:
\begin{equation}
H_J=J_1 \sum_{\langle ij\rangle}S_i\cdot S_j+J_2 \sum_{\langle ij\rangle^{\prime}}S_i\cdot S_j.
\end{equation}
Here $\langle ij\rangle$ and $\langle ij\rangle^{\prime}$ represent the summation over the nearest and next-nearest neighbors, respectively. In the present work, we set $J_1:J_2=1:0.9$.

The bare spin susceptibility determined from the tight-banding part can be calculated as,
\begin{equation}
\chi_0({\bf q},\omega)=\frac{1}{N}\sum_{\bf k}\frac{f(\varepsilon_{{\bf k}+{\bf q}})-f(\varepsilon_{{\bf k}})}{\omega-(\varepsilon_{{\bf k}+{\bf q}}-\varepsilon_{{\bf k}})+i\Gamma},
\end{equation}
where $f(x)$ is the Fermi distribution function.

The correction of the spin fluctuation induced by the spin-spin interaction $H_J$ may be included in the random phase approximation (RPA),
\begin{equation}
\chi({\bf q},\omega)=\frac{\chi_0({\bf q},\omega)}{1+J_{\bf q}\chi_0({\bf q},\omega)},
\end{equation}
where $J_{\bf q}=J_1 (\cos k_x+\cos k_y)+2J_2 \cos k_x \cos k_y$ is the Fourier factor of the spin coupling term $H_J$.

Considering that the SC pairing is meditated by the spin fluctuation, we may write the linearized eliashberg's equation,
\begin{equation}
\lambda \Delta({\bf k})=-\sum_{\bf k^{\prime}}V({\bf k}-{\bf k^{\prime}})\frac{\tanh(\beta \varepsilon_{\bf k^{\prime}}/2)}{2\varepsilon_{\bf k^{\prime}}}\Delta({\bf k^{\prime}}),
\end{equation}
where $\beta=1/T$. We consider the spin-fluctuation as the
effective pairing potential $V({\bf q})=g^2\chi({\bf q},0)$. Since we here address the pairing symmetry, we focus only on the zero-energy spin susceptibility, which should produce qualitatively correct results for the pairing symmetry as usual~\cite{kuro,xsye}.

With the SC order parameter, the bare spin susceptibility in the SC state can be expressed as,
\begin{eqnarray}
\chi_0({\bf q},\omega)&=&\frac{1}{N}\sum_{\bf k} \{
\frac{1}{2}[1+\frac{\varepsilon_{\bf k}\varepsilon_{{\bf k}+{\bf q}}+\Delta_{\bf k}\Delta_{{\bf k}+{\bf q}}}{E_{\bf k}E_{{\bf k}+{\bf q}}}]
\frac{f(E_{{\bf k}+{\bf q}})-f(E_{{\bf k}})}{\omega-E_{{\bf k}+{\bf q}}+E_{{\bf k}}+i\Gamma}\nonumber \\
&&+\frac{1}{4}[1-\frac{\varepsilon_{\bf k}\varepsilon_{{\bf k}+{\bf q}}+\Delta_{\bf k}\Delta_{{\bf k}+{\bf q}}}{E_{\bf k}E_{{\bf k}+{\bf q}}}]
\frac{1-f(E_{{\bf k}-{\bf q}})-f(E_{{\bf k}})}{\omega+E_{{\bf k}+{\bf q}}+E_{{\bf k}}+i\Gamma}\nonumber \\
&&+\frac{1}{4}[1-\frac{\varepsilon_{\bf k}\varepsilon_{{\bf k}+{\bf q}}+\Delta_{\bf k}\Delta_{{\bf k}+{\bf q}}}{E_{\bf k}E_{{\bf k}+{\bf q}}}]
\frac{f(E_{{\bf k}+{\bf q}})+f(E_{{\bf k}})-1}{\omega-E_{{\bf k}+{\bf q}}-E_{{\bf k}}+i\Gamma}\},
\end{eqnarray}
where $E_{\bf k}$ is the quasiparticle energy in the SC state with $E_{\bf k}=\sqrt{\varepsilon_{\bf k}^2+\Delta_{\bf k}^2}$. The renormalized
spin susceptibility in the framework of RPA can be obtained by Eq.(4).

\begin{figure}
\centering
  \includegraphics[width=7cm]{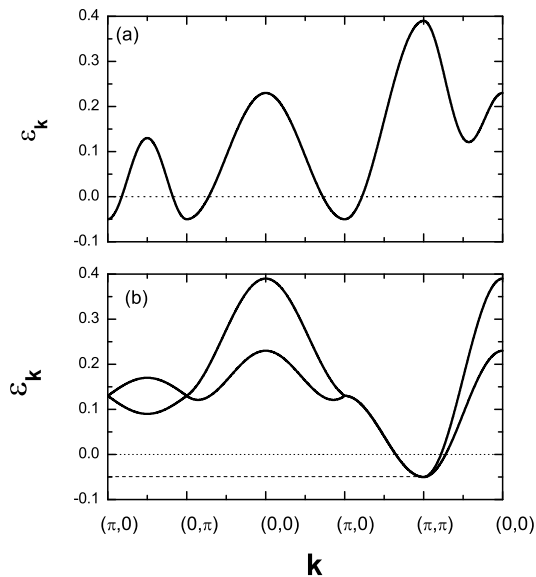}
\caption{(a) The band structure along different momentum cut in the extended BZ. (b) A replot of the band structure in the folded BZ with considering one unit cell including two Fe ions. }
\end{figure}

\section{results and discussions}
\subsection{Band structure}

Based on the five-band model, there are three bands below the Fermi energy and one band above it~\cite{yu}.
It is reasonable to concentrate on the band that crosses the FS, which plays an essential role in  superconductivity.
Considering that the electron filling per Fe site is $6+\delta$, the three filled bands contribute six electrons per site
 and thus the electron filling for the band that crosses the Fermi energy is $\delta$.  These features are crucially captured by the the present single-band model. We depict in  Fig.1(a) the band dispersion
with the hopping constants $t$ and $t^{\prime}$ being set as $0.02$ eV and $-0.045$ eV, respectively. The chemical potential is controlled
by the doping density $\delta$ with $\delta=0.2$. Considering one unit cell including two Fe irons, we plot the band dispersion in the folded brillouin zone in Fig.1(b). The low energy band dispersion is qualitatively consistent with the five-orbital model and LDA calculation~\cite{yu,jxzhu}.
While here the band width is much smaller than those obtained from five-orbital model. Actually, the five-orbital is usually fitted from the LDA calculation, which should generate qualitatively correct FS while usually the band width is larger due to neglecting the electron correlations.
In the present work, a smaller renormalized hoping constant is used and the obtained band width and doping density are consistent with the ARPES experimental results~\cite{zhang,mou,zhao,xpw,qian}.

\subsection{Pairing symmetry}
The pairing symmetry is one of the most important
issues, which can provide us the information of the pairing mechanism.  While this is still under debate for A$_x$Fe$_{2-y}$Se$_2$. Experimentally it has been revealed that the order parameter is nodeless and nearly isotropic along the FS pockets~\cite{zhang,mou,zhao,xpw}.
Since no phase-sensitive measurement has so far been reported,  both $s$-wave and nodeless $d$-wave symmetries may be able to reproduce the experimental results.
Theoretically, some predicted that the order parameter should have $d$-wave pairing~\cite{wang,tdas,mai} and some proposed the
$s$-wave symmetry~\cite{hu,mazi}. There are also some theories suggested that the pairing symmetry is either $d$-wave or $s$-wave, or even a mixture of them~\cite{sai,yzhou,yu}.

\begin{figure}
\centering
  \includegraphics[width=7cm]{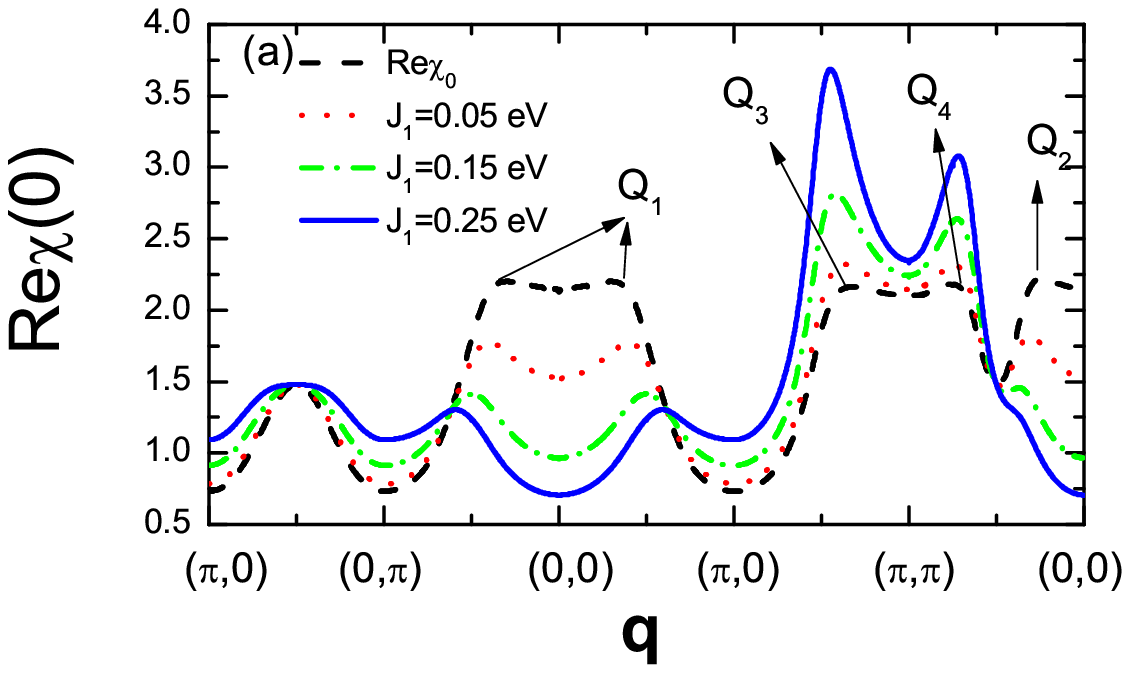}
   \includegraphics[width=7cm]{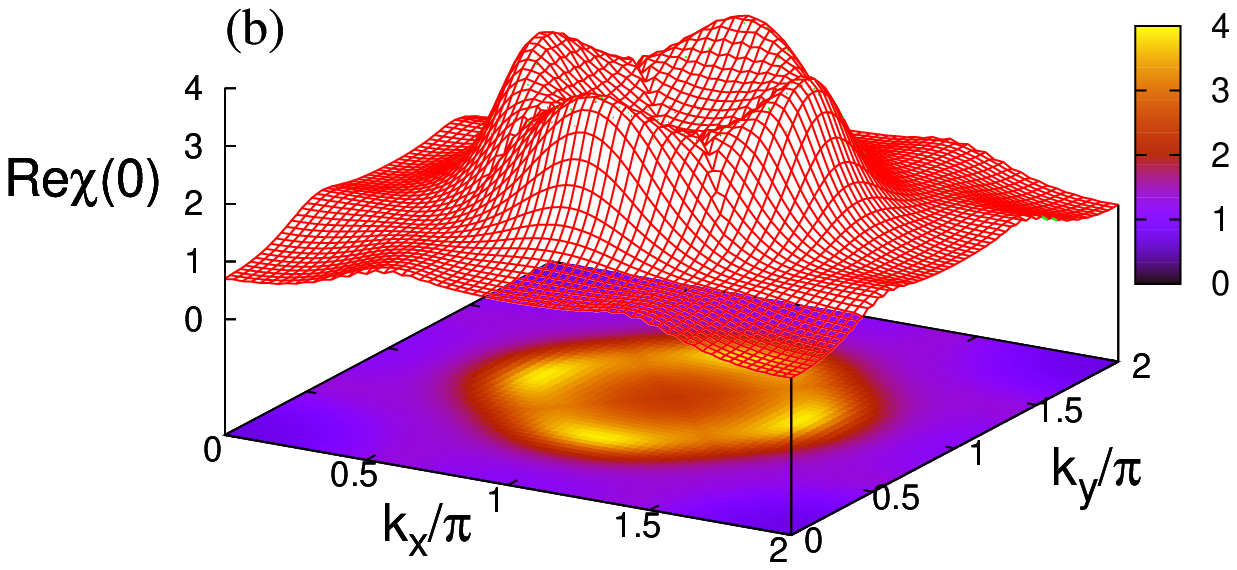}
\caption{(a) The bare and renormalized zero energy spin susceptibility with different $J_1$ along different momentum cut.
(b) The intensity plot of the renormalized spin susceptibility $\chi({\bf q},\omega=0)$ with $J_1=0.25$ ev.   }
\end{figure}

We now study the pairing symmetry based on the spin dynamics scenario. In the RPA framework, the spin susceptibility are mainly determined from the two factors:  one is the bare spin susceptibility $\chi_0$ evaluated from the tight-banding term and the other is the RPA factor $1+J_{\bf q}\chi_0$ closely related to the local spin interaction.
The behavior of the bare spin susceptibility is usually determined by the FS topology, with the maximum spin excitation occurring at the nesting wave vector. For the renormalized one, the pattern of $J_{\bf q}$, representing the spin interaction strength in the Fourier space, plays an important role.
While at present
the parameters $J_i$ in $J_{\bf q}$ are difficult to be determined theoretically.
They
usually deviate from the actually value of interchange integral constants.
In cuprates it is estimated roughly from the AF critical point~\cite{lee} or the spin resonance frequency~\cite{tzhou}. In the present work, we consider several values of $J_1$ to explore the renormalized spin susceptibility.

The bare and renormalized spin susceptibility along the two-dimensional cut are plotted in Fig.2(a). The intensity plot of the renormalized spin susceptibility in the whole BZ is plotted in Fig.2(b). As is seen, the bare spin susceptibility shows two broad features around the momentums $(0,0)$ and $(\pi,\pi)$. Here the maximum spin susceptibilities at four wave vectors ${\bf Q_{1-4}}$  are revealed.
In the RPA approach, the spin excitations around $(0,0)$ are suppressed and those around $(\pi,\pi)$ enhanced, respectively. As $J_1$ increases to 0.25 eV, the maximum spin excitation appears at the momentum ${\bf Q_3}=(\pi,\pi/2)$. The $(\pi,\pi/2)$-spin excitation can be seen more clearly from Fig.2(b). As displayed, four peaks appear at the wave vector $(\pi,\pi/2)$ and its symmetric points. This result is qualitatively consistent with the recent INS experiments~\cite{par,frie}. A further comparison between our theoretical results with INS experiments will be presented in Sec.III(C).

We here look into the pairing symmetry from Eq.(5), namely, the temperature $T$ is decreased until the maximum eigenvalue $\lambda=1$ is obtained, with the SC gap being the
eigenvector for the maximum eigenvalue. The intensity plot of the SC gap is displayed in Fig.3(a). Obviously, the pairing symmetry is of $d_{x^2-y^2}$-wave. We have checked numerically that this result is rather robust to the reasonable parameter change. The SC gap along the FS for the $d$-wave pairing symmetry $\Delta=\Delta_0/2(\cos k_x-\cos k_y)$ is plotted in Fig.3(b).
As is seen, the SC gap along one FS sheet is nearly isotropic with the difference less than 10\%, being  consistent with the ARPES experiments~\cite{zhang,mou,zhao,xpw}. In addition, it was reported in ARPES experiments that the gap is small or vanishing for the smaller pockets around the $\Gamma$ point~\cite{zhang,mou}, being also qualitatively consistent with the $d_{x^2-y^2}$-wave. However, it was indicated
in Refs.~\cite{mou,xpw} that there exists a larger FS pockets around $\Gamma$ with the gap being nearly isotropic, which seems to contradict with the $d$-wave symmetry. While we think that the larger FS around $\Gamma$ may come from the band folding effect: dominant features of the pocket and the gap may be folded from those around $M$ point. If this is the case, our results for the gap symmetry agree qualitatively with the ARPES experiments.

\begin{figure}
\centering
  \includegraphics[width=8cm]{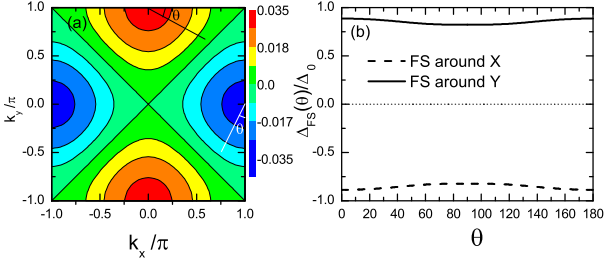}
\caption{(a) The gap function from the self-consistent calculation with the coupling strength $g=0.1$ eV. (b) The gap along the FS sheets with the dashed and solid lines are around the X and Y points, respectively, with $\theta$ denoted in panel (a).}
\end{figure}

The above results of the spin fluctuation can be addressed soundly based on the fermiology picture.
The FS is plotted in Fig.4. The FS sheets around $X$ and its symmetric points are clearly seen. We denoted the nesting wave vectors (${\bf Q_{1-4}}$) in Fig.4~\cite{note}. As seen, all of the maximum bare spin excitations shown in Fig.2a are related to the FS nesting. Based on the fermiology and Eq.(3), it is rather clear that the spin susceptibility reaches its largest value at ${\bf Q_i}$ because $\varepsilon_{{\bf k}+{\bf Q_i}}-\varepsilon_{{\bf k}}$ is vanishingly small.

We now elucidate the origin of the $d$-wave symmetry.
At the RPA approach, the spin excitations at ${\bf Q_{1,2}}$ from intra-pocket FS nestings are suppressed and those at ${\bf Q_{3,4}}$ from the inter-pocket ones enhanced. Thus the inter-pocket scattering should play major role to achieve superconductivity.
The pairing potential $V$ contributed by the spin fluctuation is largest at the wave vector ${\bf Q}={\bf Q_{3,4}}$. The factor ${\tanh(\beta \varepsilon_{\bf k^{\prime}}/2)}/{2\varepsilon_{\bf k^{\prime}}}$ in Eq.(5) is positive for any $\varepsilon_{\bf k^{\prime}}$ and largest at $\varepsilon_{\bf k^{\prime}}=0$, which
means that the pairing near the FS is important.
The gap function at or near the FS should satisfy the condition $\Delta_{\bf k}=-\Delta_{{\bf k}+{\bf Q}}$ according to Eq.(5). As shown in Fig.4, if ${\bf k}$
belongs to one sheet of FS, then ${\bf k}+{\bf Q}$ should be near the other neighboring sheet of FS. For the $d_{x^2-y^2}$-symmetry, as displayed in Fig.4(b), the SC gaps have the same magnitudes and different signs along the two neighboring FS sheets. The condition $\Delta_{\bf k}=-\Delta_{{\bf k}+{\bf Q}}$ is satisfied approximately. In this sense, we give an intuitive understanding of the pairing symmetry in this material.

\begin{figure}
\centering
  \includegraphics[width=6.2cm]{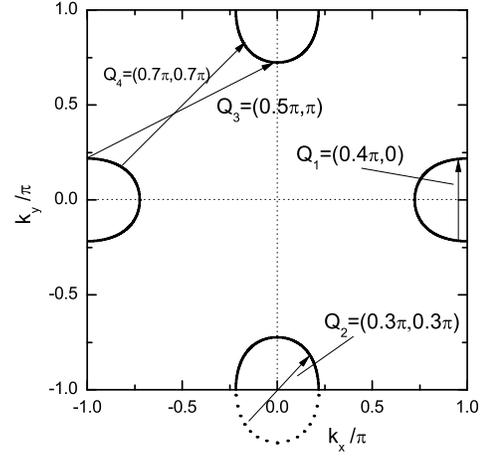}
\caption{(a) The normal state FS with the arrows denoting the different nesting vectors. }
\end{figure}

\subsection{Spin excitations in the SC state}

Now let us study the spin excitations in the SC state. Here the $d$-wave SC order parameter
$\Delta_{\bf k}=\Delta_0/2(\cos k_x-\cos k_y)$ with the maximum gap $\Delta_0=0.01$ eV being considered.
The imaginary parts of the spin susceptibility Im$\chi({\bf q},\omega)$ as a function of the wave vector ${\bf q}$ are plotted in Fig.5. One can see clearly the maximum spin excitation near the wave vector ${\bf Q}=(\pi,\pi/2)$ [with the exact peak center at $(\pi,0.515\pi)$] at $\omega=0.014$ eV. As the frequency increases to $\omega=0.018$ meV, the intensity of the spin excitation is enhanced. The maximum excitation persists at the wave vector ${\bf Q}$ [The exact peak center is at $(\pi,0.535\pi)$]. The spin excitations near the wave vector $(\pi,\pi/2)$ is well consistent with the experimental results revealed by the INS experiments~\cite{par,frie}. Moreover, the slight shift of the peak center as the frequency increases and the weak dispersion behavior are also consistent with the experimental results~\cite{frie}.

\begin{figure}
\centering
  \includegraphics[width=6.2cm]{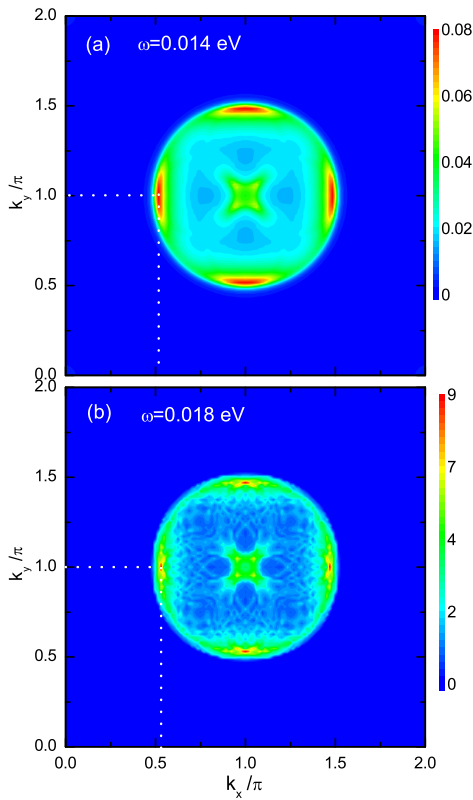}
\caption{The intensity plots of the imaginary
part of the spin susceptibility Im$\chi$
as a function of the momentum
with the energies $0.014$ eV and $0,018$ eV, respectively.}
\end{figure}

In the following, we study the frequency dependence of the  spin excitations at the wave vector ${\bf Q}=(\pi,\pi/2)$.
Here three kinds of pairing symmetry, namely, the $d_{x^2-y^2}$-wave pairing symmetry with $\Delta_{\bf k}=\Delta_0/2(\cos k_x-\cos k_y)$, $s_{x^2y^2}$-pairing symmetry with $\Delta_{\bf k}=\Delta_0\cos k_x\cos k_y$, and isotropic $s$-wave pairing symmetries with $\Delta_{\bf k}=\Delta_0$, are considered. As is known, all of the three pairings would generate isotropic SC magnitude along the FS.
The imaginary parts of spin susceptibility Im$\chi$ as a function of frequency in the normal state and SC state are presented in Fig. 6(a).
We observe clearly that for the $d$-wave pairing symmetry, the spin susceptibility at the frequency $0.018$ eV is enhanced significantly, indicating the spin resonance for this energy. For the $s_{x^2y^2}$ and isotropic $s$-wave pairing symmetry, the low energy spin susceptibilities are always less than those in the normal state. Thus there is no resonant spin excitation for these two symmetries. This significant difference between the $d$-wave symmetry and $s$-wave symmetry is interesting and this feature
can be used to determine the pairing symmetry through comparing with the experiments.

The above difference of the spin excitations can be understood through analyzing the coherence factor $C$ in Eq.(6) with
$C=1-\frac{\varepsilon_{\bf k}\varepsilon_{{\bf k}+{\bf q}}+\Delta_{\bf k}\Delta_{{\bf k}+{\bf q}}}{E_{\bf k}E_{{\bf k}+{\bf q}}}$.
The spin excitation is enhanced as $\Delta_{\bf k}\Delta_{{\bf k}+{\bf Q}}<0$ (with ${\bf k}$ and ${{\bf k}+\bf Q}$ being the momentum close to the FS). This condition can only be satisfied for the $d$-wave pairing symmetry, which provides a natural explanation for the resonant spin excitation.
The origin of the spin resonance can be clarified further in the framework of RPA. The bare spin susceptibility and the PRA factor with the $d$-wave pairing are plotted in Fig.6(b). As seen, due to the presence of the SC gap, the imaginary part of the bare spin susceptibility approaches to zero at low energies, ascribed to the spin gap. At
the edge of the spin gap (near $2\Delta_0$), it has a
steplike rise. In the mean time, the
real part of the bare spin susceptibility Re$\chi_0$ develops a
sharp structure and reaches the maximum at this frequency. Thus the RPA factor reaches the minimum (or sometimes equals to zero corresponding to a strong resonant state at the frequency $\omega<2\Delta_0$) at this frequency, which plays the major role for the appearance of the spin resonance state.

\begin{figure}
\centering
  \includegraphics[width=6.2cm]{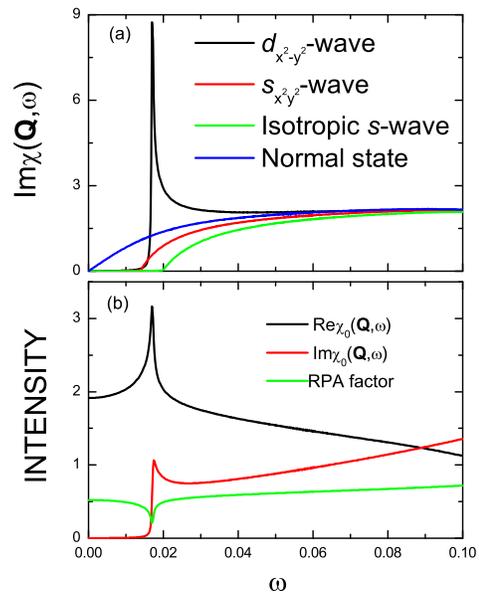}
\caption{(a) The imaginary
parts of the spin susceptibility versus the energy
for the wave vector ${\bf Q}=(\pi,\pi/2)$ in the normal state and SC states with different symmetries, respectively.
(b)The bare spin susceptibilities
versus the energy. The black
line denotes the real part and the red line
imaginary part, respectively. The green line is
the real part of the RPA factor.
}
\end{figure}

Recently the spin resonance at the wave vector ${\bf Q}=(\pi,\pi/2)$ and the frequency $0.014$ meV is observed by the INS experiments~\cite{par,frie,miaoyin}. Our numerical results for the $d$-wave pairing symmetry are qualitatively consistent with the experimental results, while those for the $s$-wave pairing symmetry contradict obviously with the experiments. In this sense, we have provided likely an indication that supports the $d$-wave pairing symmetry in this family of SC materials.

\section{summary}

In summary, we have established an effective single-band model, which captures likely the essential low-energy physics in the A$_x$Fe$_{2-y}$Se$_2$ material.
Several puzzled properties have been explained satisfactorily based on this minimum model.
 An intriguing spin excitation with the wave vector $(\pi,\pi/2)$ has been revealed.
 In addition, from the fermiology analysis, we have developed a coherent picture for the spin excitations and the unconventional $d_{x^2-y^2}$ pairing symmetry. The spin excitations in the SC states are studied and the spin resonance at $(\pi,\pi/2)$ is revealed for the $d_{x^2-y^2}$ symmetry. While there is no resonant spin excitations for the $s$-wave symmetry. Thus we have provided likely an indication that supports the $d$-wave symmetry in  A$_x$Fe$_{2-y}$Se$_2$ materials. All of our results are in qualitative agreement with the ARPES and INS experiments. In addition,
 the simple effective model presented here is quite useful and promising
for exploring rich but non-trivial physics in iron-selenide systems.

\begin{acknowledgements}
This work was supported by the NSFC under the Grant
No. 11004105, the RGC of Hong Kong under the No.
HKU7055/09P and a CRF of Hong Kong.
\end{acknowledgements}


\begin{thebibliography}{99}
\bibitem{note} The nesting is defined as the parallelism of the tangent line to FS arcs.
\end{thebibliography}
\end{document}